\documentstyle[aps]{revtex}
\begin{document}
\title{Spin orientation of two-dimensional electron gas under
intraband optical pumping}
\author{S.A.~Tarasenko and E.L.~Ivchenko}
\address{A.F.~Ioffe Physico-Technical Institute, Russian Academy of
Sciences, 194021 St.~Petersburg, Russia}
\maketitle
\begin{abstract}
The theory of spin orientation of two-dimensional (2D) electron gas
has been developed for intrasubband indirect optical transitions.
The monopolar optical orientation of electrons in the conduction band
is caused by the indirect scattering with virtual intermediate states
in the valence band and allowance for
selection rules for interband transitions. The considered mechanism
of optical orientation is shown to be in an inherent relation with
the special Elliot-Yafet mechanism of electron spin relaxation induced
by virtual interband scattering.
\end{abstract}
\section{Introduction}
In recent years spin-dependent phenomena in low-dimensional structures
attract a great attention. One of the problems of common interest is
a possibility and methods of the spin polarization in 
systems with a two-dimensional electron gas. The conventional way
to achieve the spin polarization experimentally is the 
optical orientation of electron spins~\cite{oo}. Under optical
excitation with the circularly polarized light the direct transitions
from the valence to the conduction band are allowed only
if the angular momentum is changed by $\pm 1$. These selection
rules lead to the spin orientation of optically excited electrons, with
the sign and the degree of polarization depending on the light helicity.

Up to now the theoretical consideration of the optical spin orientation
was focused on direct interband transitions \cite{oo} and,
partially, on direct intersubband transitions in the complicated valence
band $\Gamma_8$ of a zinc-blende lattice semiconductor \cite{dan}. Here
we report for the first time a theory of optical spin orientation
under the indirect intrasubband excitation of electron gas by circularly
polarized light and give an estimate of the spin generation rate.
Note that the ``monopolar" optical orientation under consideration
can serve as a model of spin injection because the only type of carriers,
electrons, is involved.

In addition, the short-range Elliot-Yafet mechanism of electron spin
relaxation considered previously only for bulk
semiconductors~(\cite{oo}, ch.~3)
is extended to quantum-well structures. This spin-relaxation
mechanism is shown to be governed by the same interband matrix elements
of the electron-phonon or electron-defect interaction as those which
govern the indirect intrasubband optical orientation.
\section{Monopolar optical orientation of electron gas}
The light absorption under intrasubband excitation in quantum wells
(Drude-like absorption) is possible only if it is assisted by
a phonon or a static imperfection in order to satisfy simultaneously both
the energy and momentum conservation laws.
Theoretically, these indirect optical transitions with initial and final
states in the same conduction subband $n$ are described by
second-order processes with virtual intermediate states. The compound matrix
element for the indirect optical transition has the standard form
\begin{equation}\label{matrelem}
M_{ns'{\bbox k}' \leftarrow ns{\bbox k}}=\sum_{\nu}
\left( \frac{ V_{ns'{\bbox k}',\,\nu{\bbox k}}\,R_{\nu{\bbox k},ns{\bbox k}} }
{E_{\nu{\bbox k}}-E_{n{\bbox k}}-\hbar\omega} +
\frac{ R_{ns'{\bbox k}',\,\nu{\bbox k}'}\,V_{\nu{\bbox k}',ns{\bbox k}} }
{E_{\nu{\bbox k}'}-E_{n{\bbox k}}} \right) \;,
\end{equation}
where $E_{n, {\bbox k}}$, $E_{n, {\bbox k}'}$ and $E_{\nu}$ are the electron
energies in the initial $|n,s, {\bbox k} \rangle$, final
$|n, s', {\bbox k}' \rangle$ and intermediate $|\nu \rangle$ states,
$s$ is the spin index, ${\bbox k}$
is the electron wavevector, $R$ is the electron-photon matrix element, 
$V$ denotes the matrix element of interaction which allows the momentum
transfer, e.g. scattering from impurities or phonon-assisted
scattering. In the latter case the photon energy $\hbar \omega$ is
assumed to exceed the energy of an involved phonon.

The most probable processes are the transitions with intermediate states
in the same subband (Fig.~1). This is the channel that determines the light
absorption. However such transitions conserve the electron spin and, hence,
{\it do not} contribute to the optical orientation.

In order to obtain the spin orientation of electron gas
under intrasubband optical transitions one should take into account the
virtual processes with intermediate states in the valence band, namely
in the heavy, light and spin-split hole subbands.
We assume that the carriers occupy the lowest electron subband $e1$.
Then in the case of the flat QW with infinite barriers,
the interband optical transitions between the $e1$ and $\nu$ states
are allowed only for the hole subbands $\nu = hh1, lh1$ and $so1$.
Fig.~2 demonstrates schematically the spin orientation under the
$\sigma^{+}$ normal-incidence excitation. In this case only the light and
spin-split hole subbands contribute to the optical orientation.
Because of the selection rules for interband matrix element,
the electron transitions with the spin reversal $|e1,-1/2 \rangle
\rightarrow |e1,+1/2 \rangle$ occur via the intermediate states
$|lh1,\pm 1/2 \rangle$ and $|so1, \pm 1/2 \rangle$,
while the inverse processes $|e1,+1/2 \rangle \rightarrow |e1,-1/2 \rangle$
are forbidden. While examining the optical orientation under oblique
incidence, the heavy hole subband should also be taken into account as
an intermediate state.

We consider the conduction electrons to form a nondegenerate 2D gas
at the temperature $T$ or a degenerate 2D gas with the Fermi energy
$\varepsilon_F$, the interaction $V$ to describe the electron scattering
by bulk acoustic phonons. Then an expression for the spin generation rate
at $\hbar\omega \gg k_B T$ or $\varepsilon_F$ ($k_B$ is Boltzmann's constant)
has the form
\begin{equation}  \label{spingen}
\dot{{\bbox S}} = \frac16 \frac{\Xi^2_{cv}}{\Xi^2_c}
\frac{\Delta^2_{so}}{E_g(E_g+\Delta_{so})(3E_g+2\Delta_{so})}
\left[ {\bbox o}_{\parallel} + \frac{a}{3} \sqrt{\frac{2 m^* \omega}
{\hbar}} {\bbox o}_{z} \right] \eta \, I_0 \, P_{circ} \:.
\end{equation}
Here $\Xi_{cv}$ and $\Xi_c$ denote the interband and intraband
deformation potential constants, $\Delta_{so}$ and $E_g$
stand for the energies of the spin-splitting and the band gap,
$a$ is the width of the quantum well, $m^*$ is the electron effective
mass, $I_0$ and $P_{circ}$ are the light intensity and the
degree of circularly polarization inside the sample, ${\bbox o}_{\parallel}$
and ${\bbox o}_z$ are the in-plane and $z$ components of the unit
vector {\bbox o} in the light propagation direction. The factor $\eta$ in
Eq.~(\ref{spingen}) is the fraction of the light energy flux absorbed in
the QW under phonon-assisted intrasubband optical transitions calculated
for intermediate states in the same conduction subband, it is given by
\begin{equation}
\eta=\frac{3 \pi \alpha}{n_{\omega}}
\left(\frac{\Xi_c}{\hbar\omega}\right)^2
\frac{k_B T}{\rho a v_s^2} \, N_e   \:,
\end{equation}
where $\alpha$ and $n_{\omega}$ are the fine structure constant
($\approx 1/137$) and the refraction index of the medium,
$\rho$ is the crystal density, $v_s$ is the sound velocity, and
$N_e$ is the 2D carrier concentration.

An estimation for typical GaAs/AlGaAs structures shows that,
at the comparable light intensities, the spin generation rate under
intrasubband phonon-assisted optical transitions is by a factor of
$10^{-4} \div 10^{-5}$ smaller than that under interband excitation.
However, consideration of other mechanisms of electron scattering,
e.g. by impurity-assisted coupling, can essentially increase this ratio.
\section{Spin relaxation mechanism induced by interband scattering}
Besides the monopolar optical orientation, the virtual interband scattering
described by the constant $\Xi_{cv}$ can lead to a short-range
mechanism of spin relaxation of the 2D electron gas.
Microscopically this mechanism is connected with
the ${\bbox kp}$-induced admixture of the valence band states,
$\Gamma_8$ and $\Gamma_7$, into the wave function of the conduction band
$\Gamma_6$ and the phonon- or defect-assisted interband coupling of
these states. For bulk semiconductors this short-range Elliot-Yafet
mechanism of electron spin relaxation was considered by Pikus and
Titkov~(see~\cite{oo}, ch.~3).

The spin relaxation time due to the mechanism under consideration
can be calculated using the spin-flip matrix element (\ref{matrelem})
where $\nu$ are the valence band states $\nu = hh1, lh1$ and $so1$
and, in the matrix element
$R_{\nu{\bbox k},ns{\bbox k}} = - (e/c m_0) {\bbox A} {\bbox p}_{\nu, ns}$,
the vector $(- e/c) {\bbox A}$ is replaced by the in-plane momentum
$\hbar {\bbox k}$. Here $e$ is the electron charge, $c$ is the light
velocity in vacuum and {\bbox A} is the amplitude of the vector potential
of the electro-magnetic wave.

Assuming the interband coupling $V$ to be caused by scattering
on bulk acoustic phonons one can derive the relaxation rates for
the in-plane and $z$ electron spin components which, after the averaging
over the Boltzmann distribution, take the form
\begin{eqnarray}\label{taus}
\frac{1}{\tau_{s \parallel}} &=& \frac16 \frac{\Xi^2_{cv}}{\Xi^2_c}
\frac{\Delta^2_{so} \, k_B T}{E_g(E_g+\Delta_{so})(3E_g+2\Delta_{so})}
\, \frac{1}{\tau_p}  \:, \\
\frac{1}{\tau_{s z}} &=& \frac{2\sqrt{\pi}a}{3}
\sqrt{\frac{2m^*}{\hbar^2}k_B T}
\frac{1}{\tau_{s \parallel}} \:,
\end{eqnarray}
where the momentum relaxation time determined by the acoustic
phonon-assisted scattering is given by
\begin{equation}
\frac{1}{\tau_p}=\frac32 \frac{m^* \, \Xi_{c}^2}{\rho a v_s^2 \hbar^3}
\, k_B T \:.
\end{equation}
One can see that, in case of the scattering by acoustic phonons, the spin
relaxation of the 2D electron gas governed by the short-range Elliot-Yafet
mechanism is very anisotropic, $\tau_{s \parallel} \ll \tau_{s z}$.
\subsection*{Acknowledgements}
This work was supported by the Russian Ministry of Science,
the RFFI (projects 00-02-16997 and 01-02-17528), the INTAS (project
99-00015), and by the Science Programme of the Presidium of RAS "Quantum
low-dimensional structures".

\end{document}